\documentclass[conference]{IEEEtran}
\IEEEoverridecommandlockouts

\usepackage{amsmath,amssymb,amsfonts}
\usepackage{algorithmic}
\usepackage{graphicx}
\usepackage{textcomp}
\usepackage{caption}
\usepackage{svg}
\usepackage[caption=false,font=normalsize,labelfont=sf,textfont=sf]{subfig}
\usepackage{xcolor}
\usepackage{stfloats}
\usepackage[per-mode=symbol,retain-explicit-plus,separate-uncertainty=true,detect-all=true]{siunitx}
\def\defaultDigitGrouping{5}
\usepackage{balance}
\usepackage[sorting=none]{biblatex}
\AtBeginBibliography{\footnotesize}

\begin{document}

\title{X-HEEP: An Open-Source, Configurable\\and Extendible RISC-V Platform\\for TinyAI Applications}

\author{
\IEEEauthorblockN{Simone Machetti\IEEEauthorrefmark{1}, 
Pasquale Davide Schiavone\IEEEauthorrefmark{1}, 
Giovanni Ansaloni\IEEEauthorrefmark{1}, 
Miguel Peón-Quirós\IEEEauthorrefmark{7}, 
David Atienza\IEEEauthorrefmark{1}} \\
\IEEEauthorblockA{\IEEEauthorrefmark{1}Embedded Systems Laboratory, EPFL, Lausanne, Switzerland}
\IEEEauthorblockA{\IEEEauthorrefmark{7}EcoCloud, EPFL, Lausanne, Switzerland} \\
\IEEEauthorblockA{simone.machetti@epfl.ch, davide.schiavone@epfl.ch, giovanni.ansaloni@epfl.ch,\\ miguel.peon@epfl.ch, david.atienza@epfl.ch}
}

\maketitle

\IEEEpubid{\makebox[\columnwidth][l]{\hspace{-4.6cm}\footnotesize
979-8-3315-3477-6/25/\$31.00~\copyright~2025 IEEE}}

\begin{abstract}

    In this work, we present \mbox{X-HEEP}, an open-source, configurable, and extendible RISC-V platform for ultra-low-power edge applications (TinyAI). \mbox{X-HEEP} features the eXtendible Accelerator InterFace (XAIF), which enables seamless integration of accelerators with varying requirements along with an extensive internal configuration of cores, memory, bus, and peripherals. Moreover, it supports various development flows, including FPGA prototyping, ASIC implementation, and mixed SystemC-RTL modeling, enabling efficient exploration and optimization. Implemented in TSMC’s \SI{65}{\nano\meter} CMOS technology (\SI{300}{\mega\hertz}, \SI{0.8}{\volt}), \mbox{X-HEEP} achieves a minimal footprint of only \SI{0.15}{\milli\meter\squared} and consumes just \SI{29}{\micro\watt} of leakage power. 
    As a demonstrator of the configurability and low overhead of \mbox{X-HEEP} as a host platform, we present a study integrating it with near-memory accelerators targeting early-exit dynamic network applications, achieving up to \SI{7.3}{\times} performance speedup and \SI{3.6}{\times} energy improvement on the resulting heterogeneous system compared to CPU-only execution.

\end{abstract}

\IEEEpubidadjcol


\section{Introduction}

    The rapid rise of edge computing is driven by the increasing demand for real-time, privacy-preserving data processing near the source. Applications such as autonomous vehicles, wearable health monitors, and smart industrial systems increasingly require low-latency and energy-efficient computation. However, as edge deployments become more common, they encounter significant performance and power constraints, particularly on resource-limited devices with stringent area and thermal budgets.

    Heterogeneous architectures have emerged as a promising approach to address these challenges. These architectures can deliver higher performance and energy efficiency by combining ultra-low-power host processors for control tasks with domain-specific accelerators for workloads such as machine learning, image processing, healthcare, and cryptography. However, each accelerator introduces unique memory, area, performance, and power requirements, which makes it essential to have a highly configurable host platform that can adapt to diverse system demands~\cite{co-design_vision}. Whereas commercial solutions often restrict hardware exploration through proprietary constraints and licensing fees, open-source platforms offer greater flexibility and control over intellectual property (IP), enabling a wider range of research and product innovation.

    Propelled by the momentum of the RISC-V revolution, a variety of open-source heterogeneous platforms have emerged in recent years~\cite{pulp, cgra, nmc, asip}. However, many of these lack the configurability necessary to adapt internal components, such as core, memory, and bus, support external accelerators, and implement dynamic power management strategies. Consequently, developers who aim to integrate custom accelerators frequently face substantial design overhead, requiring extensive manual modifications to hardware and software stacks.

    To overcome these limitations, we present \mbox{X-HEEP}\footnote{X-HEEP repository: \url{https://github.com/esl-epfl/x-heep}.}~\cite{x-heep, x-heep_1}, an open-source and highly configurable platform specifically designed for ultra-low-power edge applications (TinyAI). Built on the RISC-V instruction set architecture (ISA), \mbox{X-HEEP} features the eXtendible Accelerator InterFace (XAIF), which enables straightforward and flexible integration of tightly-coupled accelerators with different area, power, and performance constraints. Furthermore, \mbox{X-HEEP} provides extensive internal configurability, enabling fine-grained optimizations based on domain-specific requirements. Through its modular design, \mbox{X-HEEP} allows users to explore various hardware design trade-offs, supporting development workflows across FPGA prototyping, ASIC implementation, and mixed SystemC-RTL modeling. 

    To showcase the capabilities of the \mbox{X-HEEP} framework, we present an heterogeneous System-on-Chip (SoC) featuring near-memory acceleration~\cite{nmc}, synthesized in TSMC’s \SI{65}{\nano\meter} CMOS technology and operating at \SI{300}{\mega\hertz} and \SI{0.8}{\volt}. A dynamic network benchmark is then used to evaluate the design, estimating the area and leakage overhead introduced by \mbox{X-HEEP}, as well as the performance and energy improvements achieved by the complete system.

    The remainder of this paper is structured as follows. Section~2 examines state-of-the-art host platforms. Section~3 describes the \mbox{X-HEEP} platform. Section~4 summarizes relevant integrations based on \mbox{X-HEEP}. Sections~5 and Section~6 introduce and evaluate the above-mentioned \mbox{X-HEEP}-based SoC. Finally, Section~7 concludes the paper.

\section{State-of-the-art}

    A widely used open-source hardware and ultra-low-power platform is PULPissimo~\cite{pulpissimo}, a single-core design that supports CV32E20 or CV32E40P cores. Its integration with various accelerators (including the PULP cluster~\cite{pulp}) has demonstrated high energy efficiency. However, PULPissimo does not offer a natively configurable and rich interface to integrate domain-specific accelerators. Still, it offers a \SI{32}{\bit} AXI slave and a \SI{64}{\bit} AXI master interface, limiting the bandwidth that external accelerators can exploit. It also lacks support for external interrupts and power control, as well as the possibility to configure the memory, bus, and peripherals natively, reducing its flexibility.

    On the other hand, Cheshire~\cite{cheshire} addresses some of these limitations by offering configurable external master/slave ports, last-level cache (LLC) size, and peripherals. Although this provides additional flexibility, Cheshire targets higher-performance systems than X-HEEP, employing a Linux-capable CPU and consuming up to \SI{300}{\milli\watt}, making it unsuitable for scenarios requiring a few tens of milliwatts. Moreover, Cheshire still lacks support for external interrupts, power control, a configurable core, bus, and memory.

    For higher performance applications, BlackParrot~\cite{blackparrot} offers a tile-based Linux-capable design that supports heterogeneous accelerator integration. It enables the composition of \SI{64}{\bit} cores, L2 cache slices, I/O modules, and accelerators. Despite this flexibility, BlackParrot does not provide configurability for core, bus, or memory. In addition, it lacks common peripherals such as I2C, GPIO, DMA, timers, and a power manager.

    Security-centric platforms such as OpenTitan~\cite{opentitan} take a different approach by emphasizing cryptographic robustness for ultra-low-power use cases. Although it includes a single CV32E20 core and a comprehensive peripheral set, it does not natively support external accelerator plug-ins, requiring developers to modify the RTL to add accelerators manually. In addition, it lacks configurability in core, memory, and power management.

    In contrast, Chipyard~\cite{chipyard} provides a more flexible framework based on the Chisel hardware description language derived from the Rocket Chip generator. It supports configurable core, memory, and peripherals, offering high adaptability. However, Chipyard requires accelerators to be integrated directly into Chisel, imposing a steep learning curve and a non-industrial standard flow. In addition, it lacks native power management capabilities, necessitating a great effort to implement energy optimization strategies.

    Although Chipyard is primarily geared toward ASIC development, LiteX~\cite{litex} serves as a configurable system generator that mainly targets FPGA-based platforms. It offers extensive customization of core, memory, and external master/slave ports. However, unlike PULP-based solutions, LiteX does not support ASIC workflows, which complicates silicon implementations. Like BlackParrot and OpenTitan, LiteX lacks built-in support for external interrupts and power control, making effective power management more challenging.

    Lastly, ESP~\cite{esp} employs a network-on-chip (NoC) interconnect for heterogeneous accelerator integration. While this approach is structured and scalable for high-performance edge applications, its coarse-grained tile-based design is less suited to ultra-low-power edge scenarios requiring tens of milliwatts, but rather targets systems with a few watts as power profiles.

    Taken together, these limitations of the analyzed platforms reveal a gap for a solution that delivers enhanced configurability, robust power management, and seamless accelerator integration tailored explicitly to TinyAI.

\section{X-HEEP Platform}

    \begin{figure*}[t]
        \centering
        \includegraphics[width=0.8\textwidth]{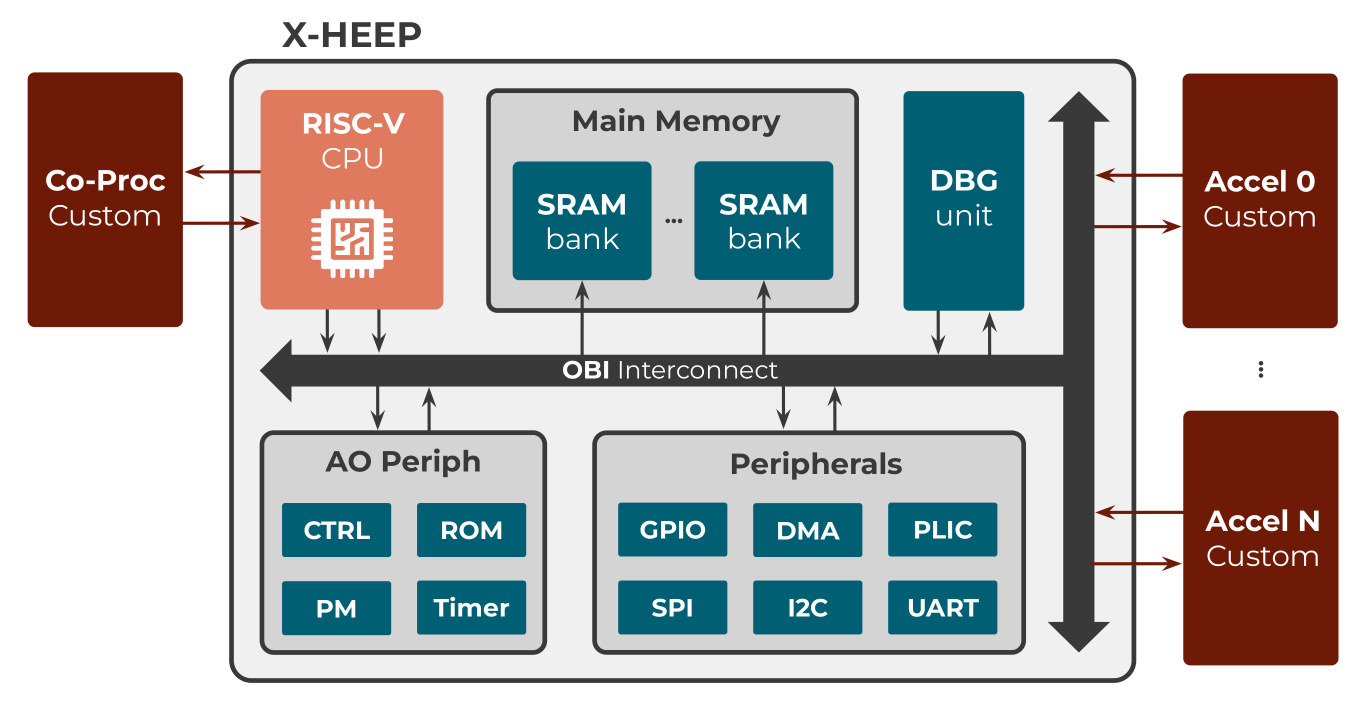}
        \caption{\textit{X-HEEP} architecture with integrated accelerators and co-processor.}
        \label{fig:x-heep}
    \end{figure*}

    In this section, we provide a detailed description of the key features that distinguish \mbox{X-HEEP}~\cite{x-heep_1, x-heep_2} regarding configurability, extensibility, software stack, and implementations.

    \subsection{Hardware Components}

        Figure~\ref{fig:x-heep} illustrates the architecture of \mbox{X-HEEP}, highlighting its primary building blocks: a configurable RISC-V core, flexible buses, configurable memory, multiple peripheral domains, and extensible accelerator interfaces. SystemVerilog parameters and templates drive the automatic generation of tailored RTL designs, enabling developers to easily fine-tune architectural features without relying on new higher-level languages.

        \mbox{X-HEEP} builds upon widely recognized open-source hardware IPs to maximize compatibility with existing ecosystems and to streamline both hardware and software development. The CPU cores are sourced from the OpenHW Group CORE-V family, which is renowned for its robust verification infrastructure and proven silicon implementations. Additional IPs, such as bus, peripherals, and memory, are mainly adapted from the PULP project, while other peripheral components are drawn from the OpenTitan initiative~\cite{opentitan}. Complementing these, \mbox{X-HEEP} integrates several custom-designed modules, including an efficient power manager, a fast interrupt controller, and a smart multi-channel DMA engine, supporting both 1D and 2D transactions. All major IPs are written in SystemVerilog to ensure straightforward integration into standard industrial EDA toolchains.

        Designers can choose between the CV32E20, CV32E40X, CV32E40P, and CV32E40PX cores~\cite{cve2, cv32e40p} to balance power consumption and computational performance according to the target use case. CV32E40X and CV32E40PX offer the CORE-V-XIF interface~\cite{core-v-xif} to integrate co-processors for RISC-V ISA extensions.

        The memory is designed to be highly configurable, allowing the total size, addressing scheme, and number of banks to be tuned. If supported by the memories provider, each bank can be independently switched on and off via clock- and power-gating switches or set in retentive mode, substantially reducing power when idle. This feature supports energy-efficient designs without compromising memory availability.

        Communication across the platform is managed by buses based on the open bus interface (OBI)~\cite{obi}. The topology can be customized: A one-at-a-time configuration minimizes resource usage for simpler systems, whereas a fully connected crossbar maximizes parallelism and throughput for performance-intensive designs.

        The main peripheral domain includes essential system components, such as a platform-level interrupt controller (PLIC), timers, and commonly used communication interfaces, including GPIO, I2C, I2S, and SPI. To optimize for power, the peripheral domain can be clock- or power-gated when not in use. Meanwhile, each peripheral can be selectively removed to save area. A separate always-on domain ensures the availability of essential services, housing modules such as the SoC controller, boot ROM, power manager, fast interrupt controller, DMA, and a minimal set of communication peripherals. The integrated power manager implements clock gating, power gating, and memory retention strategies.

        The XAIF further enhances the system extensibility by providing a standardized accelerator interface. It bundles configurable OBI connections (slave and master), DMA extensions, interrupts, and power management control signals. This design empowers accelerators not only to access shared memory resources efficiently, but also to autonomously manage their life cycle through dynamic power control. When combined with the CV32E40X or CV32E40PX cores, XAIF includes CORE-V-XIF support, allowing custom accelerators to extend the RISC-V instruction set transparently.

    \subsection{Software Components}

        \mbox{X-HEEP} complements its hardware flexibility with a comprehensive and extensible software stack, covering configuration, development, and deployment workflows via a hardware abstraction layer (HAL) and support for FreeRTOS.

        The simulation and synthesis flows are managed by the FuseSoC framework~\cite{fusesoc}. Open-source tools, such as Verilator, and commercial tools can be exploited for simulation. For synthesis flows, the Synopsys Design Compiler and Vivado tools are used for ASIC and FPGA targets, respectively.

        A wide range of application examples complete the repository to help developers use the \mbox{X-HEEP} framework. Applications can be compiled with either the RISC-V GCC or LLVM toolchain.

    \subsection{Implementations}

        \mbox{X-HEEP} supports multiple implementation targets, catering to a wide range of evaluation and deployment scenarios, from early-stage validation to final silicon tape-out.

        For ASIC implementations, the platform includes a pad ring and pad controller generator, as well as technology-independent templates for memory wrapper and standard cells. These components facilitate clean integration with foundry-specific technology libraries and simplify adaptation across different CMOS technology nodes.

        FPGA deployments are supported in two primary modes: standalone and Linux-based. In the standalone configuration, \mbox{X-HEEP} is synthesized onto the programmable logic (PL) region of FPGAs. At the time of writing,  support is provided for the Pynq-Z2, the ZCU104, and the Nexys A7 boards. Standard board connectivity enables straightforward communication between the platform and external devices.

        In the Linux-based configuration, \mbox{X-HEEP} runs on the PL side, while Linux operates on the processing system (PS) side. This heterogeneous setup enables interaction between Linux user-space applications and the hardware platform through a custom Python API. This environment, named the FPGA EMUlation Platform (FEMU)~\cite{femu}\footnote{\textit{FEMU} repositories: \url{https://github.com/esl-epfl/x-heep-femu-sdk} and \url{https://github.com/esl-epfl/x-heep-femu}.}, significantly accelerates development workflows by allowing developers to test and interact with hardware modules through high-level software without requiring low-level FPGA reprogramming or hardware description knowledge.

\section{X-HEEP-based Systems}

    This section highlights representative works that use \mbox{X-HEEP} to integrate custom accelerators in diverse domains, including machine learning, signal processing, arithmetic, and cryptography. Rather than providing an exhaustive survey, we highlight below examples that demonstrate how \mbox{X-HEEP} enables state-of-the-art research for TinyAI systems, offering enhanced configurability and extensibility for integrating accelerators across a wide range of application scenarios.

    \subsection{Memory-like Accelerators}

        Several accelerators have been integrated to offload compute-intensive tasks by writing to external accelerators' memories. Blade~\cite{blade}, integrated into the HEEPocrates chip~\cite{heepocrates}, leverages in-memory computing (IMC) techniques to deliver up to \SI{4.8}{\times} energy savings.  NM-Carus and NM-Caesar~\cite{nmc} embed programmable vector units directly within SRAM banks, minimizing data movement overhead and achieving up to \SI{53.9}{\times} acceleration. Based on this approach, the ARCANE~\cite{arcane} architecture aggregates multiple NM-Carus units with an LLC, reaching up to \SI{84}{\times} speedup with a modest area overhead.
        
        Another example is the HEEPstor framework~\cite{systolic_array}, which incorporates a hybrid-quantized systolic array to accelerate matrix multiplication (GEMM) for TinyML workloads, achieving up to \SI{4.5}{\times} speedup over CPU execution on Fashion-MNIST. 

    \subsection{Master Accelerators}

        Moving beyond pure slave interfaces, several accelerators adopt a master-slave model, allowing them to autonomously access the system memory. Within the HEEPocrates chip~\cite{heepocrates}, the OpenEdge coarse-grained reconfigurable array (CGRA)~\cite{cgra, bio_apps} enhances flexibility for programmable feature extraction, achieving up to \SI{4.9}{\times} energy savings. Further, extending reconfigurability, the STRELA architecture~\cite{strela} introduces a streaming elastic CGRA capable of efficiently handling both data- and control-driven workloads, reaching up to \SI{18.6}{\times} speedup.

        For massively parallel TinyAI applications, the edge Graphic Processing Unit (e-GPU)~\cite{e-gpu} couples lightweight programmability with energy efficiency, achieving up to \SI{7}{\times} acceleration and \SI{45}{\percent} energy improvement, while maintaining an area overhead below \SI{1.3}{\times}. 

        To meet bandwidth demands, accelerators often add many ports to the bus, increasing congestion and impacting performance and power. For accelerators with DMA-compatible access patterns, elastic DMA interfaces can collect data without additional master ports. An example is the Im2Col accelerator presented in ~\cite{dma}, which leverages the X-HEEP DMA multi-channel and 2D addressing scheme to perform runtime data-layout transformations, delivering up to \SI{6.1}{\times} speedup for data reshaping operations.

    \subsection{Coprocessors}

        Tightly coupled ISA extensions can be implemented via dedicated coprocessors, extending the CPU pipeline through the CORE-V-X-IF interface without requiring modifications to the CPU RTL. For example, the Quadrilatero coprocessor~\cite{quadrilatero} integrates a matrix coprocessor, achieving up to \SI{3.87}{\times} speedup and \SI{15}{\percent} energy savings. However, for addressing low-precision arithmetic applications using floating-point, the Coprosit accelerator~\cite{percival} leads to area savings of \SI{38}{\percent} and energy reductions of \SI{54}{\percent} relative to standard IEEE 754 floating-point units. Finally, in the cryptographic domain, the ATHOS accelerator~\cite{athos} offloads post-quantum cryptographic algorithms, providing \SI{7.74}{\times} and \SI{4.12}{\times} performance improvements for the CRYSTALS-Kyber and CRYSTALS-Dilithium security workloads, respectively, with an area overhead of only \SI{1.47}{\times}.

\section{Experimental Setup}

    Exemplifying the capabilities of the \mbox{X-HEEP} approach, this section presents a study on the performance and energy benefits of near-memory acceleration for dynamic networks implemented on the \mbox{X-HEEP} platform. We first introduce the selected application domain and benchmark models, then describe the integrated accelerator and its deployment, and finally detail the methodology used for performance and energy consumption analysis.

    Dynamic networks are an emerging class of algorithms that adapt computation at runtime by terminating inference once a sufficient confidence level is reached. This enables significant energy savings and latency reductions in TinyAI systems. Among these, early-exit strategies offer an effective trade-off between computational cost and prediction accuracy.

    This study evaluates two such adaptive models, a transformer~\cite{transf_app} and a convolutional neural network (CNN)~\cite{cnn_app}, for seizure detection in bio-signal processing~\cite{ace}, a domain characterized by highly unbalanced data distributions. Each model is augmented with a single early-exit point after its first major processing stage (i.e., the first encoder layer for the transformer and the first convolutional block for the CNN). These exit points are selected through a parameter sweep to minimize energy consumption while preserving accuracy.

    Both models are retrained under varying early-exit loss weights (0.001–0.1), entropy thresholds (0.1–0.5), and initialization strategies. Pretrained backbones consistently yield better early-exit performance. The optimization goal is to maximize the early-exit rate while maintaining an acceptable degradation in F1-score.

    In the final configuration, the transformer model uses an early-exit weight of 0.1 and an entropy threshold of 0.45, achieving a \SI{73}{\percent} early-exit rate with a reduction in F1-score from 0.6223 to 0.53. The CNN model, configured with a weight of 0.01 and a threshold of 0.35, achieves an even higher early-exit rate of \SI{82}{\percent}, with the F1 score decreasing from 0.57 to 0.49.

    To evaluate the benefits of the software strategies analyzed, we integrate the near-memory accelerator NM-Carus~\cite{nmc} with \mbox{X-HEEP}. The resulting heterogeneous system is synthesized using TSMC’s \SI{65}{\nano\meter} CMOS technology and operates at \SI{300}{\mega\hertz} and \SI{0.8}{\volt}. Post-synthesis simulations are used to assess performance and extract switching activity, which is then fed into power analysis tools for accurate energy estimation. The measured energy consumption is compared against a baseline CPU-only execution to quantify the improvements enabled by near-memory acceleration.

    \begin{figure}[t]
        \centering
        \includegraphics[width=0.24\textwidth]{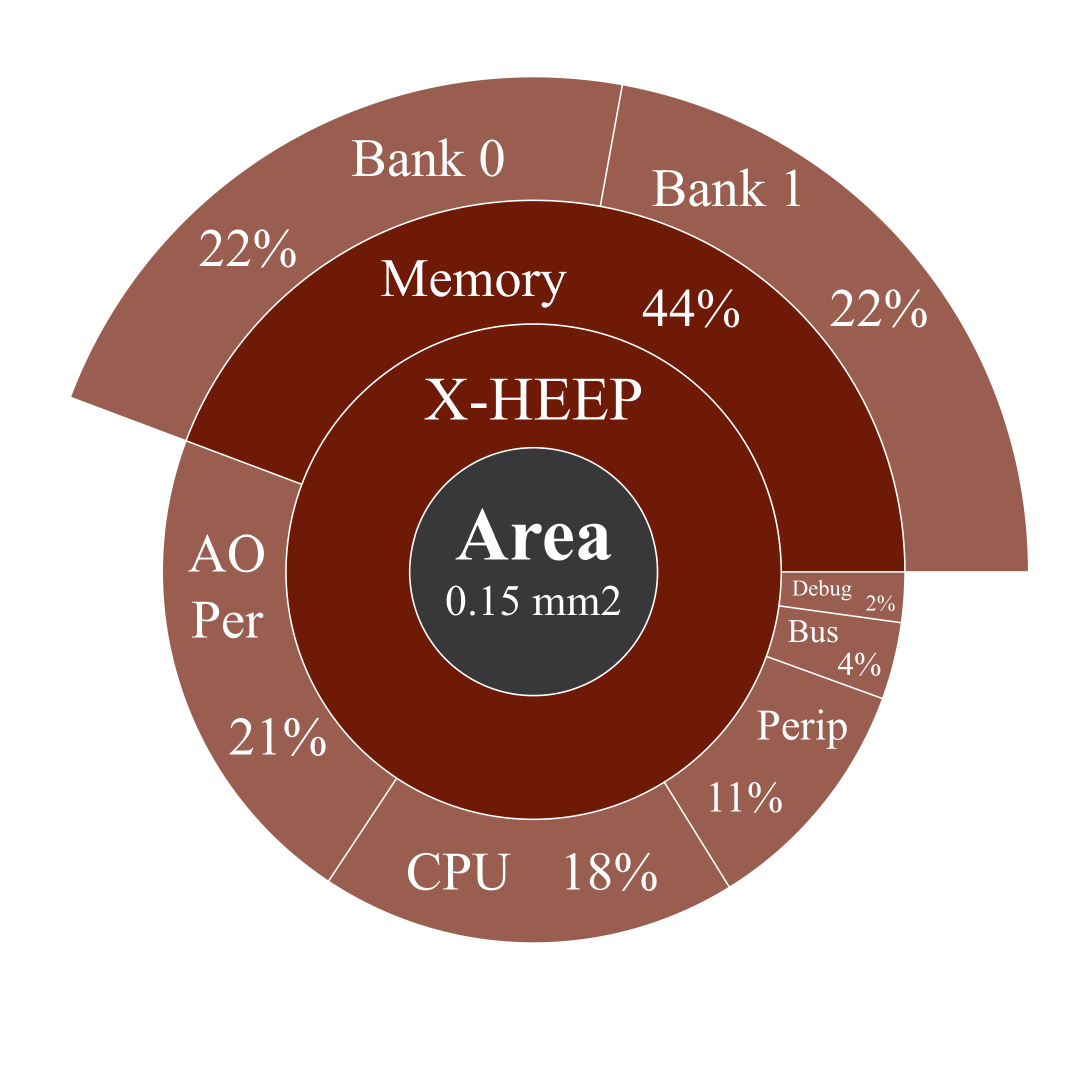}
        \includegraphics[width=0.24\textwidth]{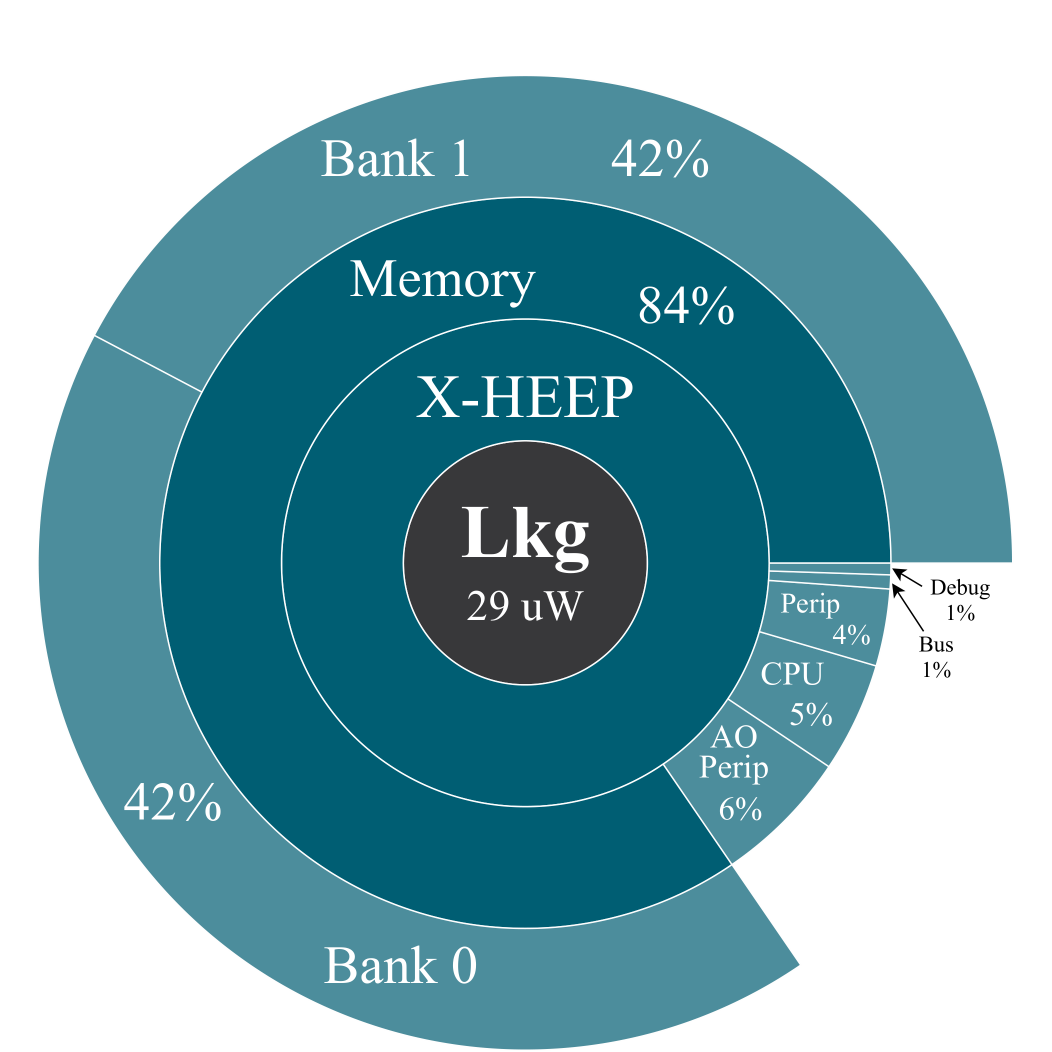}
        \caption{Area and leakage distributions among the internal components of the X-HEEP platform.}
        \label{fig:pie}
    \end{figure}

\section{Experimental Results}

    This section presents and analyzes the results obtained from our experiments. We first evaluate the area and leakage characteristics of the standalone \mbox{X-HEEP} host to assess its integration overhead. Then, we compare the complete system, including the Carus accelerator~\cite{nmc}, against the host CPU in terms of performance and energy consumption.
    
    \subsection{Static characterization of the X-HEEP host}

        Figure~\ref{fig:pie} illustrates the area distribution among the internal components of the \mbox{X-HEEP} architecture, which occupies a total area of \SI{0.15}{\milli\meter\squared}. The two memory banks dominate the design, accounting for a combined \SI{44}{\percent} of the total area, while the always-on (AO) peripheral and the standard peripheral subsystems occupy \SI{21}{\percent} and \SI{11}{\percent}, respectively. The CV32E40P CPU~\cite{cv32e40p} itself consumes only \SI{18}{\percent} of the area, and the bus and debug modules introduce minimal overheads of \SI{4}{\percent} and \SI{2}{\percent}, respectively. This breakdown highlights that the integration overhead introduced by \mbox{X-HEEP} is low, with most of the area attributed to the memory subsystems, which can be scaled according to the requirements of the target applications rather than to the control and computation logic.

        Similarly, the \mbox{X-HEEP} architecture exhibits a total leakage of \SI{29}{\micro\watt}. The memory banks once again dominate, accounting for a combined \SI{84}{\percent} of the leakage power, with each bank contributing \SI{42}{\percent}. The always-on (AO) peripheral and the standard peripheral subsystems contribute \SI{6}{\percent} and \SI{4}{\percent}, respectively. The CPU accounts for only \SI{5}{\percent} of the leakage, while the bus and debug modules each introduce a negligible \SI{2}{\percent}. As in the area distribution, this breakdown confirms that the leakage overhead introduced by \mbox{X-HEEP} remains minimal, with most static power consumption dominated by memory. In addition, the CPU, peripheral subsystems, and individual memory banks can be selectively powered down through the platform power manager when not in use, further reducing overall leakage to a minimum of \SI{3}{\micro\watt}.

    \subsection{Run-time characterization of the heterogeneous system}

\begin{figure}[t]
        \centering
        \includegraphics[width=0.49\textwidth]{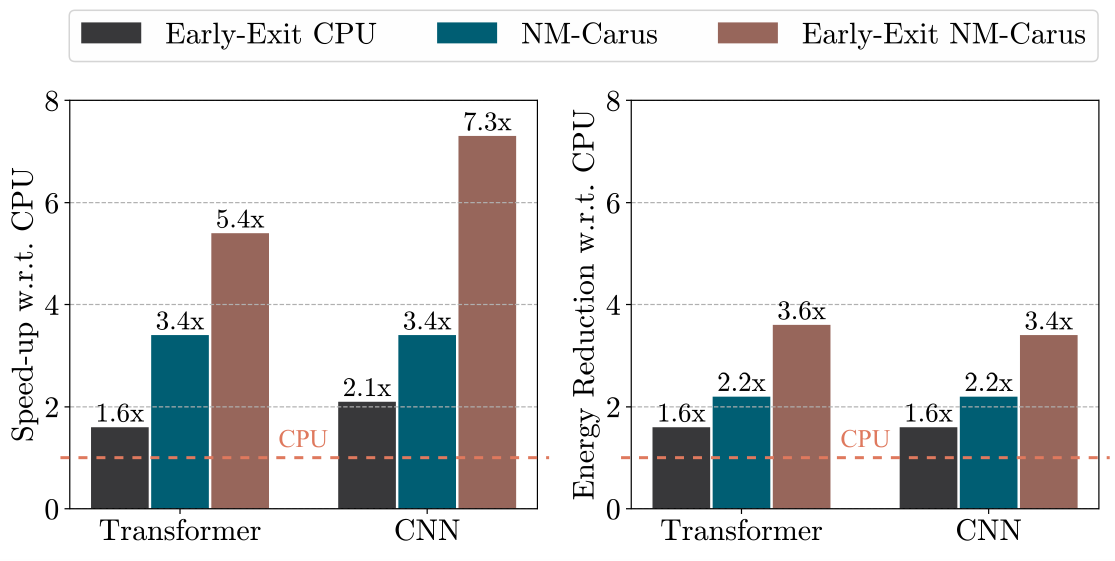}
        \caption{Performance and energy improvements obtained by running the adaptive network benchmark with early exit and on the NM-Carus accelerator compared to the baseline CPU-only execution.}
        \label{fig:dyn}
    \end{figure}

        Figure~\ref{fig:dyn} reports the performance and energy efficiency obtained when running the adaptive network benchmarks. We evaluate three configurations: (i) early-exit inference executed on the host CPU, (ii) standard inference offloaded to the NM-Carus accelerator, and (iii) early-exit inference offloaded to the NM-Carus accelerator. All results are normalized against the baseline execution on the host CPU without early exit.

        In terms of kernel-level performance, early-exit inference on the CPU achieves speed-ups of up to \SI{1.6}{\times} for the Transformer model and \SI{2.1}{\times} for the CNN model. Offloading to the NM-Carus, which targets integer arithmetic, further enhances performance, reaching up to \SI{3.4}{\times} (Transformer) and \SI{3.4}{\times} (CNN) speed-up without early exit and up to \SI{5.4}{\times} (Transformer) and \SI{7.3}{\times} (CNN) when combining early exit with near-memory acceleration.

        Similarly, energy efficiency significantly improves across all configurations. Early-exit execution on the CPU alone achieves energy gains of up to \SI{1.6}{\times} (Transformer) and \SI{1.6}{\times} (CNN). Offloading to NM-Carus provides up to \SI{2.2}{\times} (Transformer) and \SI{2.2}{\times} (CNN) improvement without early exit, and up to \SI{3.6}{\times} (Transformer) and \SI{3.4}{\times} (CNN) when combining early exit with the accelerator.

        Across all analyzed configurations, the CNN model exhibits greater power increases relative to CPU-only execution than the Transformer model. This higher power overhead limits potential energy savings, leading to slightly better overall energy efficiency for the Transformer case study.

\section{Conclusion}

    In this work, we have presented \mbox{X-HEEP}, an open-source, configurable, and extendible RISC-V platform designed to address the challenges of integrating domain-specific accelerators in TinyAI systems. Using widely adopted open source IPs, \mbox{X-HEEP} provides extensive configurability concerning types of cores, memory hierarchy, buses configuration, and supported peripherals while offering seamless accelerator integration through the XAIF interface. Its flexible architecture enables efficient trade-off explorations across both FPGA and ASIC implementations. Synthesized in TSMC’s \SI{65}{\nano\meter} CMOS technology (\SI{300}{\mega\hertz}, \SI{0.8}{\volt}) \mbox{X-HEEP} achieves a minimal footprint of only \SI{0.15}{\milli\meter\squared} and consumes just \SI{29}{\micro\watt} of leakage power. When integrated with the NM-Carus accelerator to support early-exit dynamic networks, the resulting heterogeneous system reaches up to \SI{7.3}{\times} performance speedup and \SI{3.6}{\times} energy improvement compared to CPU-only execution, demonstrating how our configurable and low-overhead host platform can support substantial gains in real-world edge scenarios.

\section{Acknowledgements}

    The authors thank the entire \mbox{X-HEEP} team for their significant contributions to creating the final platform. Furthermore, we express our gratitude to Anna Burdina for her valuable support in compiling the information to complete the experimental part of this paper.

    This work was supported in part by the Swiss State Secretariat for Education, Research, and Innovation (SERI) through the SwissChips research project, and in part by the Swiss NSF, grant no. 10.002.812: ”Edge-Companions: Hardware/Software Co-Optimization Toward Energy-Minimal Health Monitoring at the Edge”.

\printbibliography

\end{document}